 \documentclass[aps,prb,twocolumn,superscriptaddress,showpacs]{revtex4-1}

 \usepackage{graphicx,psfrag}
 \usepackage[latin1]{inputenc}
 \usepackage{natbib}
 \usepackage{amsmath, amsthm, amssymb}
 \usepackage{color}
 \usepackage{comment}
 \usepackage{hyperref}

 \newcommand{\notes}[1]{}

 \newcommand{\beq}{\begin{equation}}
 \newcommand{\eeq}{\end{equation}}
 \newcommand{\beqnn}{\begin{equation*}}
 \newcommand{\eeqnn}{\end{equation*}}
 \newcommand{\beqas}{\begin{eqnarray*}}
 \newcommand{\eeqas}{\end{eqnarray*}}
 \newcommand{\beqa}{\begin{eqnarray}}
 \newcommand{\eeqa}{\end{eqnarray}}
 
\DeclareMathOperator{\sgn}{sgn}
\newcommand{\mvec}{{\bf m}}

\newcommand{\xivec}{{\bf \xi}}
\newcommand{\pvec}{{\bf p}}

\begin{document}

\title{Micromagnetic instabilities in spin-transfer switching of perpendicular magnetic tunnel junctions}

\author{Nahuel Statuto}
\affiliation{Center for Quantum Phenomena, Department of Physics, New York University, New York, New York 10003 USA}
\author{Jamileh Beik Mohammadi}
\affiliation{Department of Physics, Loyola University New Orleans, New Orleans, LA 70118 USA}
\author{Andrew D. Kent}
\affiliation{Center for Quantum Phenomena, Department of Physics, New York University, New York, New York 10003 USA}

\date{\today}

\begin{abstract}
Micromagnetic instabilities and non-uniform magnetization states play a significant role in spin transfer induced switching of nanometer scale magnetic elements. Here we model domain wall mediated switching dynamics in perpendicularly magnetized magnetic tunnel junction nanopillars. We show that domain wall surface tension always leads to magnetization oscillations and instabilities associated with the disk shape of the junction. A collective coordinate model is developed that captures aspects of these instabilities and illustrates their physical origin. Model results are compared to those of micromagnetic simulations. The switching dynamics is found to be very sensitive to the domain wall position and phase, which characterizes the angle of the magnetization in the disk plane. This sensitivity is reduced in the presence of spin torques and the spin current needed to displace a domain wall can be far less than the threshold current for switching from a uniformly magnetized state. A prediction of this model is conductance oscillations of increasing frequency during the switching process.
\end{abstract}
\maketitle
\ifpdf
\graphicspath{{Figs/}}
\fi

\section{Introduction}
 Spin transfer torque magnetization switching has been extensively studied since it was first predicted theoretically~\cite{Slonczewski1989,Slonczewski1996,Berger1996} and demonstrated experimentally in spin-valve nanopillars~\cite{Katine2000}. The magnetic anisotropy of the free layer plays an important role in setting the switching current; materials with a uniaxial anisotropy exhibit far more efficient spin-transfer switching~\cite{Sun2000}. This has led to research on easy axis perpendicular to the plane free layers. Advances include the demonstration of spin-transfer switching in perpendicularly magnetized spin-valve nanopillars~\cite{Mangin2006,Ravelosona2006} and perpendicularly magnetized magnetic tunnel junctions~\cite{Ikeda2010,Worledge2011}. When the free layer is in the shape of a disk there is axial symmetry that simplifies the analysis of the magnetization dynamics. Thus recent research has focused on understanding the magnetization switching mechanisms in this high symmetry situation~\cite{Hahn2016,Devolder2016_w,Devolder2016_a}. Perpendicular magnetic tunnel junctions nanopillars (pMTJs) are also under intense development for applications as magnetic random access memory (MRAM)~\cite{Kent2015,Thomas2017,Slaughter2017,Jinnai2020}.  

Such junctions consist of thin ferromagnetic metallic layers, one with a magnetization free to reorient and the other with a fixed magnetization direction separated by a thin insulating barrier. The junction stable magnetic states are layers magnetized parallel (P state) or antiparallel (AP state) and have conductances that differ by a factor of 2 (or more) with CoFeB electrodes and an MgO insulating barrier~\cite{Butler2001,Yuasa2004,Parkin2004}. Current flow through the junction leads to spin-transfer torques on the free layer magnetization that can switch it between magnetic states. In the macrospin limit---where the switching between these states is by coherent spin rotation---there are analytic models that characterize the thermally activated switching~\cite{Chaves2015,Thomas2017,Mihajlovic2020} and spin-transfer driven switching~\cite{Sun2000,Adam2012,Liu2014}. 

Experiments, however, suggest that the magnetization reverses nonuniformly and the reversal process appears to be reversed domain nucleation and expansion by domain wall motion~\cite{Hahn2016,Devolder2016_a,Devolder2016_w}. This is confirmed by micromagnetic modeling, which shows that the assumption of a coherent magnetization reversal does not capture the switching dynamics above a critical diameter set by the exchange interaction strength, magnetic anisotropy and magnetization~\cite{Bouquin2018,Mohammadi2020,Volvach2020} $d_c=(16/\pi) \sqrt{A/K_\mathrm{eff}}$,
where $A$ is the exchange constant and $K_\mathrm{eff}(d) = K_p - \mu_0M^2\big[3N_{zz}(d)-1\big]/4$ is the diameter-dependent effective anisotropy. Here $K_p$ is the perpendicular magnetic anisotropy, $\mu_0$ the permability of free space, $M$ is the magnetization and $N_{zz}(d)$ is the demagnetization coefficient perpendicular to the plane of the free layer that depends on the element's diameter $d$ and its thickness. For state-of-the-art pMTJs the critical diameter $d_c$  can be just 10 to 30 nm, as the exchange constant of the thin CoFeB free layer can be much less than the bulk value~\cite{Mohammadi2019}. As the diameter increases beyond $d_c$ for fixed current overdrive $j/j_c$---the current $j$ divided by the threshold current for magnetization switching $j_c$---the switching time increases and the average magnetization is a nonmonotonic function of time~\cite{Mohammadi2020}.

\begin{figure}[t]
  \centering
   \includegraphics[width=1\columnwidth]{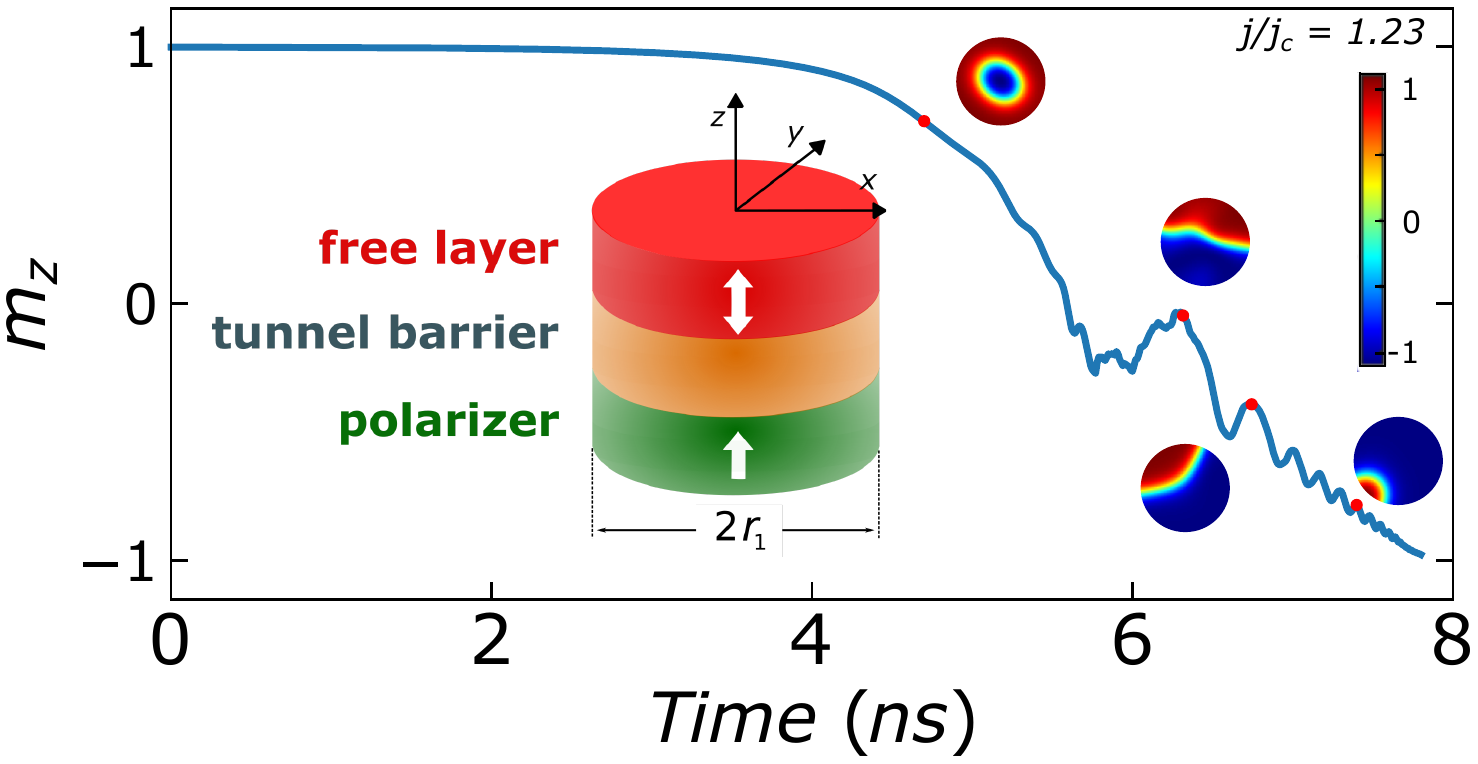}
    \caption{Time evolution of the spatially averaged normalized z-axis disk magnetization during a nucleation process with $j/j_c=1.23$ and $d/d_c=1.95$ ($r_1=15$ nm). The other material parameters can be found in Appendix~\ref{Sec:AppendixPartE}. Snapshots show the disk magnetization at different times indicated in the curve by red dots. The inset is a schematic of the tunnel junction nanopillar.}
\label{Fig:nucleation}
\end{figure}

The micromagnetic simulation in Fig.~\ref{Fig:nucleation} illustrates this reversal process. The free layer's spatially averaged perpendicular ($z$-axis) component of magnetization is plotted versus time along with magnetization images at various times in the reversal process. The reversal starts with the formation of a reversed region in the center of the free layer which then experiences a drift instability, leading to a domain wall that traverses the element~\cite{Mohammadi2020,Volvach2020}. There are then magnetization oscillations of increasing frequency as the reversed domain expands to complete the reversal. This behavior appears generic to reversed domain expansion by domain wall motion in a disk.

This paper addresses the origin of the magnetization oscillations in the switching process both with an analytic model and micromagnetic simulations. A model for the collective coordinates of a domain wall in a disk is developed and its solutions illustrated. The origin of the magnetization oscillations is shown to be Walker breakdown associated with domain wall surface tension which produces an effective force on the domain wall that varies with its position. Micromagnetic simulations are used to test the model and determine features of the collective coordinate model that are preserved as more spin degrees of freedom are considered.  We also discuss the implications of these micromagnetic instabilities on magnetic tunnel junction characteristics and propose experiments to observe this phenomena. 

The structure of the paper is as follows. Section~\ref{Sec:AM} introduces the analytic model and its solutions for magnetization relaxation from different initial states. Section~\ref{Sec:MM} presents micromagnetic simulations that are compared to the analytic model. In Sec.~\ref{Sec:SPC} the effect of a spin current is studied and in Sec.~\ref{Sec:EC} some of the experimental consequences of the model are described. The main results are then summarized. The Appendices include more details on the derivation of the analytic model (Secs.~\ref{Sec:AppendixPartA}-\ref{Sec:AppendixPartC}) and micromagnetic simulations (Secs.~\ref{Sec:AppendixPartD}-\ref{Sec:AppendixPartE}).

\section{Analytic Model}
\label{Sec:AM}


The starting point for our analysis is the Landau-Lifshitz-Gilbert (LLG) equation with a spin-torque term:
\begin{equation}
{\bf \dot{m}}=-\gamma \mu_0 {\bf m} \times {\bf H}_\text{eff} +\alpha {\bf m} \times {\bf \dot{m}}+a_I \mvec \times (\mvec \times \pvec),
\label{Eq:LLGS}
\end{equation}
where ${\bf m}$ is a unit vector in the direction of magnetization of the free layer, $\gamma$
is the gyromagnetic ratio, $\mu_0$ the permeability of free space, and ${\bf H}_\text{eff}$ is the effective field.
The effective field is the variational derivative of the energy density, $U$, with respect to the magnetization:
${\mu_0\bf H}_\text{eff}= -(1/M)\delta U/(\delta {\bf m})$, where $M$ is the magnetization of the free layer.
The second term on the right is the damping term, where $\alpha$ is the Gilbert damping constant ($\alpha \ll 1$).
The last term on the right hand side is the spin-transfer torque, $\pvec$ is the direction of the spin polarization and $a_I$ is proportional to the spin current $a_I=\hbar P j/(2eM t)$, where $j$ is the current density, $\hbar$ is the reduced Planck's constant, $P$ is the spin polarization of the current and $t$ is the thickness of the free layer.

\begin{figure}[t]
  \centering
   \includegraphics[width=1\columnwidth]{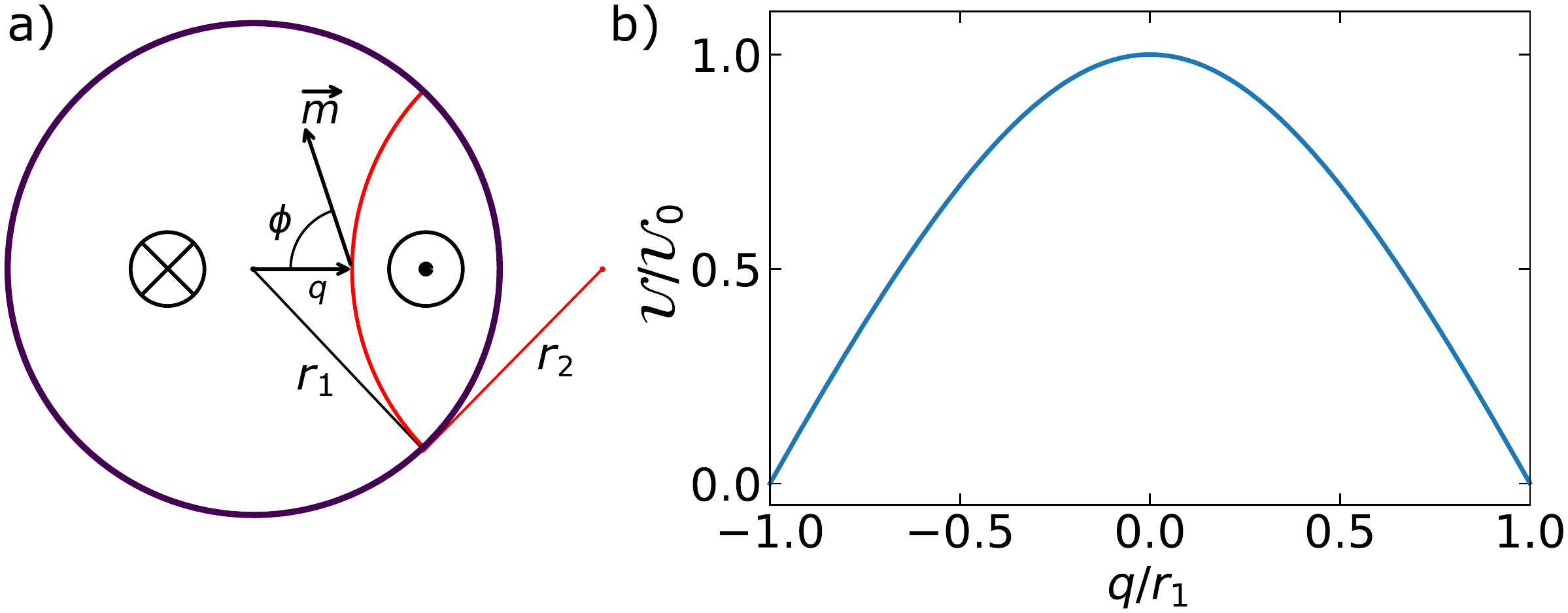}
    \caption{a) Schematic of a reversed domain in a thin disk of radius $r_1$. The domain boundary---the red line---is characterized by its position from the center of the disk $q$ and the angle of the magnetization from the boundary normal $\phi$. The curvature of the DW is represented by radius $r_2$. b) Energy of the DW as a function of position in the disk normalized to its energy $\mathcal{U}_0$ at $q=0$, its energy at the center of the disk.}
\label{Fig:schematic_energy}
\end{figure}

The magnetization texture can be written in cylindrical coordinates with the origin at the center of the arc characterizing the DW (i.e. radius $r_2$):
\begin{eqnarray}
m_z(r)&=&-\tanh((r-r_2)/\Delta), \nonumber \\ 
m_r (r)&=&\cos \phi /(\cosh((r-r_2)/\Delta), \nonumber\\
m_\theta(r)&=&\sin \phi  /(\cosh((r-r_2)/\Delta),
\label{Eq:ms}
\end{eqnarray}
where $r$ is measured from the center of this same circle. $\Delta$ characterizes the width of the domain wall; the full width of the domain wall is typically taken to be $\pi \Delta$~\cite{Hubert1998}.

The magnetic energy density is given by:
\begin{equation}
U=A \left[(\nabla m_x)^2 + (\nabla m_y)^2 + (\nabla m_z)^2\right]-K_\mathrm{eff}m_z^2,
\label{Eq:energydensity}
\end{equation}
where $A$ is the exchange constant of the material and $K_\mathrm{eff}$ is the effective perpendicular magnetic anisotropy, the difference between the perpendicular uniaxial anisotropy and demagnetizing energy (i.e dipole-dipole interactions are described by an average demagnetization field)~\cite{Chaves2015}. The first term on the right hand side of Eq.~\ref{Eq:energydensity} is thus the exchange energy and the second term is the magnetic anisotropy energy. Energy minimization gives $\Delta=2\sqrt{A/K_\mathrm{eff}}$. The total energy $\mathcal{U}$ is obtained by substituting the expressions for $\bf{m}$ in Eqs.~\ref{Eq:ms} into Eq.~\ref{Eq:energydensity} and integrating it over the volume of the disk (see Appendix~\ref{Sec:AppendixPartA}):
\begin{equation}
\mathcal{U}=\left[\sigma_\mathrm{DW}+\frac{\Delta \mu_0 H_N M}{2} \cos^2 \phi \right] t \ell,
\label{Eq:totalenergy}
\end{equation}
where $\ell$ is the length of the domain wall. The first term on the right in Eq.~\ref{Eq:energydensity}, $\sigma_\mathrm{DW}$, is the energy associated with a Bloch wall ($\phi=\pi/2$) per unit wall area, $\sigma_\mathrm{DW}=4\sqrt{AK_\mathrm{eff}}$. The next term is the added energy density associated with a N\'eel wall ($\phi=0$) and $H_N$ is the applied field normal to the DW that would transform a Bloch wall into a N\'eel wall, $H_N \approx tM/(t+\pi \Delta)$ (for $\ell \gg \Delta$). We do not include a term in the energy associated with the curvature of the DW (characterized by radius $r_2$) as this term only becomes significant when the DW curvature is of order of its width. Under these conditions the DW exits the sample and the model assumptions for the domain wall energy no longer hold (see Appendix~\ref{Sec:AppendixPartB}). The energy of the domain wall is plotted versus $q$ in Fig.~\ref{Fig:schematic_energy}(b). It is thus clear that the shape of element leads to a conservative force on the domain wall $F_q=-\partial \mathcal{U}/\partial q$ that tends to expel the DW from the disk.

\begin{figure}[t]
  \centering
  \includegraphics[width=1\columnwidth]{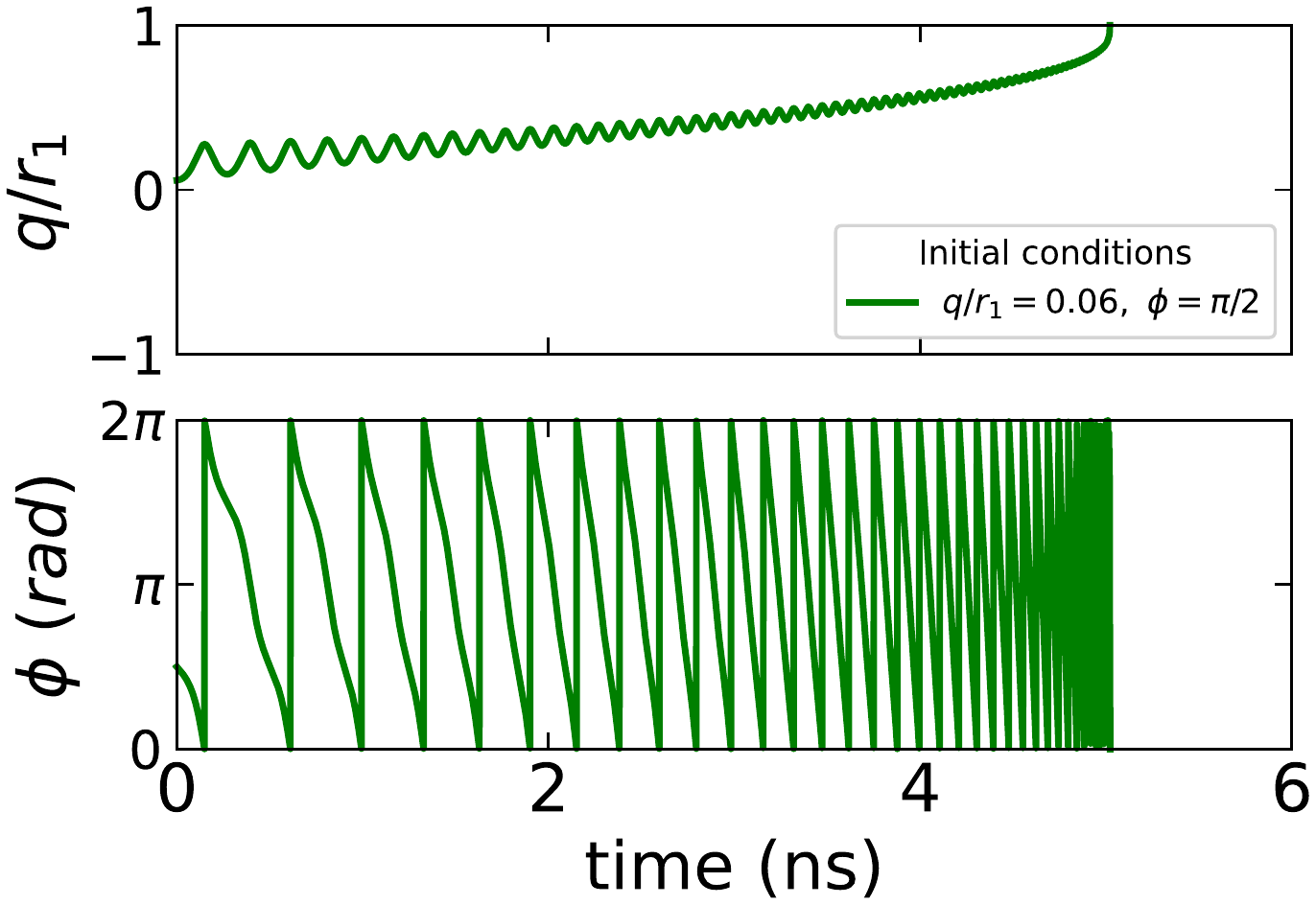}
     \caption{Domain wall dynamics in the ($q,\phi$) model starting at $t=0$ in a Bloch configuration ($\phi=\pi/2$) with $q/r_1=0.06$. The upper panel shows $q$ and the lower panel shows the phase, restricted to $[0,2\pi]$, as a function of time. There is no spin-polarized current or applied magnetic field.}
     \label{FIG:relax_bloch_phase_model}
\end{figure} 
We derive an equation for the collective coordinates and the generalized velocities of the DW in the disk from Eq.~\ref{Eq:LLGS} following the approach outlined in Ref.~\cite{Tretiakov2008} (see Appendix~\ref{Sec:AppendixPartA} for further details).
The DW coordinates ($q$,$\phi$) satisfy a Thiele equation, similar to that of a DW moving under the influence of spin-transfer torque in a nanowire~\cite{Thiaville2005,Tretiakov2008,Cucchiara2012,Devolder2016_w}:
\begin{eqnarray}
\dot{\phi}&-&\frac{\alpha}{\Delta} \dot{q}+ 
\frac{\gamma}{2t \ell M} \frac{\partial \mathcal{U}}{\partial q}=0 
\label{Eq:phi}
\\
-\dot{q}/\Delta&-&\alpha \dot{\phi}+\frac{\gamma \mu_0 H_N }{4}\sin{2\phi}-\frac{\pi a_I}{2}=0,
\label{Eq:q}
\end{eqnarray}
where an overdot indicates a derivative with respect to time. The last term in Eq.~\ref{Eq:phi} is proportional to the conservative force $F_q$ associated with the free layer's shape. It has been denoted a Laplace pressure~\cite{Zhang2018} or a domain wall tension. Again, $a_I$ is proportional to the current and the last term in Eq.~\ref{Eq:q} is the spin transfer torque on the domain wall due to the spin-polarized current from the reference layer. Eq.~\ref{Eq:phi} indicates that the Laplace pressure tends to cause spins in the domain wall to precess, while a spin torque couples directly to the domain wall displacement (Eq.~\ref{Eq:q}). The wall velocity is related to $H_N$ and the spin-transfer torque with maximum domain wall velocity (for $\phi=\pi/4$ and no spin torque) given by:
$v_\mathrm{max}=\Delta \gamma \mu_0 H_N/4$.
	
We illustrate the DW relaxation dynamics in Fig.~\ref{FIG:relax_bloch_phase_model}. A Bloch wall ($\phi=\pi/2$) is started at a position just to the right of the center of the disk, i.e. $q\gtrapprox 0$ and moves to the right (average $q$ increasing) with its position oscillating as a function of time. The domain wall phase $\phi$ runs, indicating precession of the spins in the DW.
This behavior corresponds to DW motion motion in the Walker breakdown limit, i.e. when $\dot{\phi} \neq 0$ the domain wall position oscillates and moves at an average velocity less than $v_\mathrm{max}$. In the absence of a spin current and in zero magnetic field the condition for Walker breakdown is (see Appendix~\ref{Sec:AppendixPartA}):
\begin{equation}
\frac{1}{2d\ell M}\frac{\partial U}{\partial q}\simeq \frac{\sigma_\mathrm{DW}}{2M t\ell}\frac{d\ell}{dq} >\alpha \frac{\mu_0 H_N}{4}.
\label{Eq:Walker}
\end{equation}
$(d\ell/dq)/\ell$ diverges as the domain wall reaches the element boundary $q\simeq \pm r_1$. Thus the domain wall position {\em always} oscillates in a magnetization reversal process that occurs by reversed domain expansion in a disk, with an oscillation frequency that increases as the domain wall approaches the element boundary. This is the characteristic seen in the micromagnetic modeling in Fig.~\ref{Fig:nucleation} for $m_z<0$ showing that the collective coordinate model captures elements of the physics of magnetization reversal that occurs by DW propagation.

\begin{figure}[t]
  \centering
  \includegraphics[width=1\columnwidth]{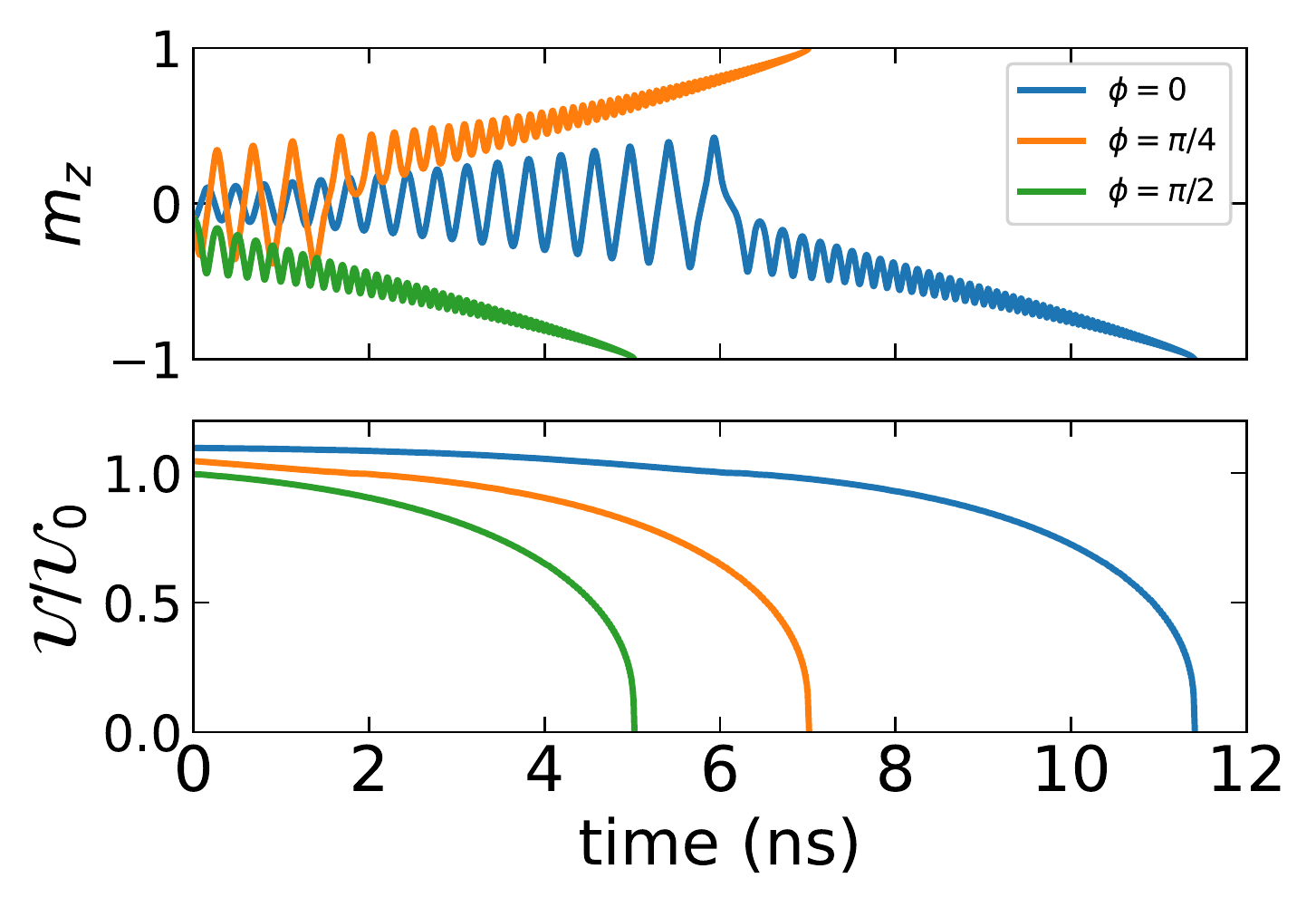}
     \caption{Domain wall dynamics in the ($q,\phi$) model starting from $q/r_1=0.06$ and different initial DW phases: N\'eel ($\phi=0$) and Bloch ($\phi=\pi/2$) and an intermediate phase, $\phi=\pi/4$. Again, the applied field and spin current are zero.}
     \label{FIG:relax_curves_model}
\end{figure}

Interestingly, the model further predicts non-trivial DW dynamics and magnetic relaxation that depends sensitively on the initial DW phase $\phi$. Figure~\ref{FIG:relax_curves_model} shows DW dynamics for the same initial position and different initial phases and plots the average z-magnetization $m_z$ and the total magnetic energy versus time. The oscillation frequency and amplitude vary significantly with time. This behavior is in contrast to that of DW motion in a nanowire~\cite{Thiaville2005}. Surprisingly, the final state changes with initial $\phi$, for $\phi=0$ and $\phi=\pi/2$ the magnetizaton relaxes to $m_z=-1$, a down magnetized domain. Whereas for $\phi=\pi/4$ the magnetization relaxes to $m_z=+1$, an up magnetized domain, even from an initial state $m_z<0$. Further, the relaxation time scale changes significantly with the phase. This illustrates the important role of the DW phase in the dynamics. 

\begin{figure}[t]
  \centering
  \includegraphics[width=1\columnwidth]{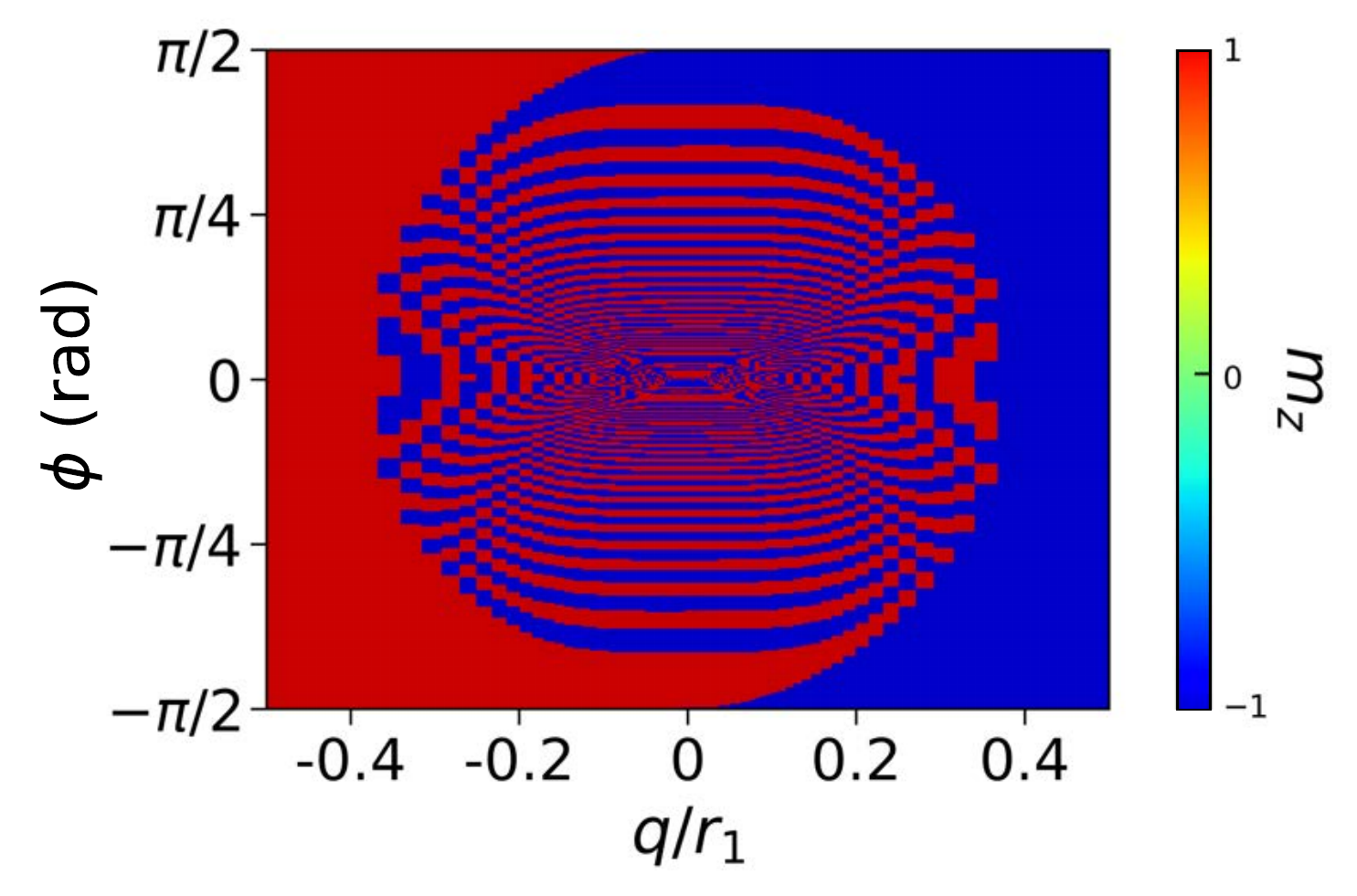}
     \caption{Relaxation phase diagram showing the final magnetization state as a function of the initial conditions for the analytic model. Red represents a final state with magnetization up and blue magnetization down.}
     \label{FIG:plus_minus_q}
\end{figure}

To explore this further a DW relaxation phase diagram was computed from the analytic model. The final state (magnetization up or magnetization down) is computed as a function of $q$ and $\phi$. The results are shown in Fig.~\ref{FIG:plus_minus_q}. The blue color represents magnetization relaxation to a down state ($m_z=-1$) and the red color represents magnetization relaxation to an up state ($m_z=+1$). The intricate pattern highlights the sensitivity to the the initial DW position and phase. It is straightforward to include the effect of an applied field which modifies the pattern as discussed in Appendix~\ref{Sec:AppendixPartC}.

\section{Micromagnetic model} 
\label{Sec:MM}
A basic question is to what extent this simple collective coordinate model with two degrees of freedom ($q,\phi$) captures the DW dynamics in a ferromagnetic disk. The full problem is much more complex and can include variations in the DW curvature (i.e. that the DW is not rigid), its width as well as non-local magnetic dipole interactions that are not considered in this simple model. For this reason, we performed micromagnetic simulations and compared them to our collective coordinate model.
 
\begin{figure}[t]
  \centering
  \includegraphics[width=1\columnwidth]{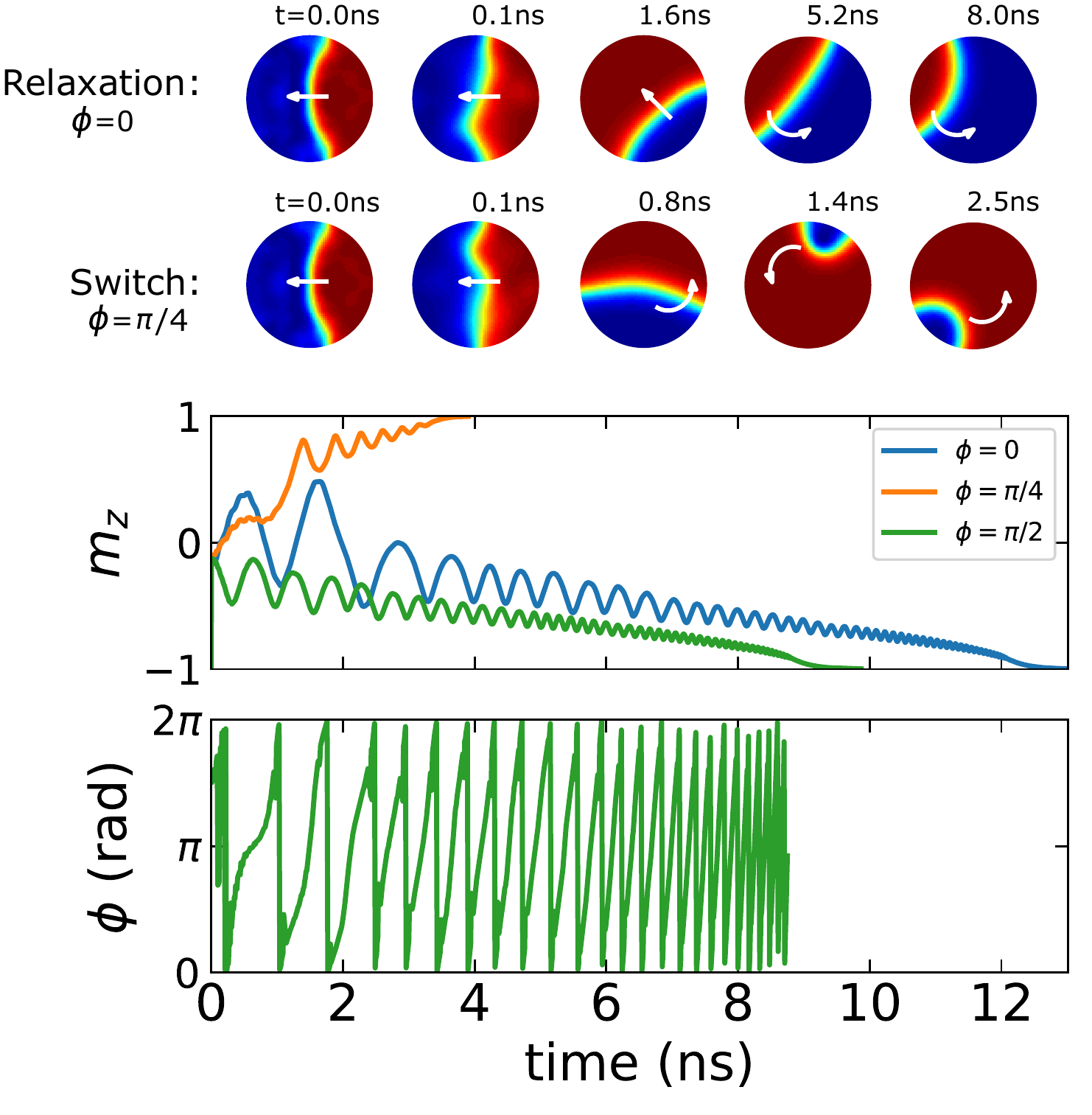}
     \caption{Time evolution of $m_z$ for the same initial DW position but different initial phases $\phi$. We simulate a CoFeB disk with radius $15$ nm and thickness $2.3$ nm with $q/r_1=0.06$ nm with no spin current or applied magnetic field. Starting from both N\'eel ($\phi=0$) and a Bloch DW ($\phi=\pi/2$) the disk relaxes to down state. Whereas for an initial phase of $\phi=\pi/4$ the magnetization relaxes into an up state for the same initial DW position. Bottom panel shows the time evolution of the Bloch DW's phase, which oscillates with an increasing frequency as the domain is expelled to the edge of the disk. Relaxation panels: Starting form a N\'eel DW ($\phi=0$) the disk relaxes to down state. Whereas in the panels labeled switch the initial phase is $\phi=\pi/4$ and relaxation is to an up state for the same initial DW position. White arrows indicate the direction of the DW displacement.}
     \label{FIG:relax_curves}
\end{figure}

Micromagnetic simulations were performed using the open-source MuMaX$^3$ code \cite{Vansteenkiste2014}. We model the magnetic disk using the same parameters as in the analytic model and the initial states were two magnetically-opposed domains with a DW at the same position as in Fig.~\ref{Fig:schematic_energy}(a). Figure~\ref{FIG:relax_curves} shows the micromagnetic simulations for the same initial conditions as those in Fig.~\ref{FIG:relax_curves_model}. The central panel shows the time evolution of the normalized disk magnetization, $m_z$, for reversed domains starting at the same position, $q/r_1=0.06$ with $m_z<0$, and at three different initial domain wall phases. For N\'eel and Bloch configurations, $\phi=0$ and $\pi/2$ respectively, the system relaxes towards the down magnetic state whereas for an intermediate angle $\phi=\pi/4$, orange curve, the up magnetic domain expands resulting in switching of the disk magnetization. Top panels show snapshots of the magnetization at different times for the initial N\'eel configuration and the initial intermediate, $\phi=\pi/4$ configuration; the top panel shows a relaxation process and the lower panel shows a case in which the magnetization switches. Moreover, the bottom panel shows the time evolution of the phase in the initial Bloch configuration, which was calculated from the magnetization vector images. The phase, as in Fig.~\ref{FIG:relax_curves_model}, oscillates from $0$ to $2\pi$ indicating that the DW spins are oscillating as a function of the time and their frequency is increasing as the domain is expelled from the edge of the disk. The same behavior is obtained no matter the initial phase: the magnetization oscillates, the magnetic domain breathes and the domain wall moves back and forward indicating that the system is in the Walker breakdown regime as seen in our analytic model. 

\begin{figure}[t]
  \centering
  \includegraphics[width=1\columnwidth]{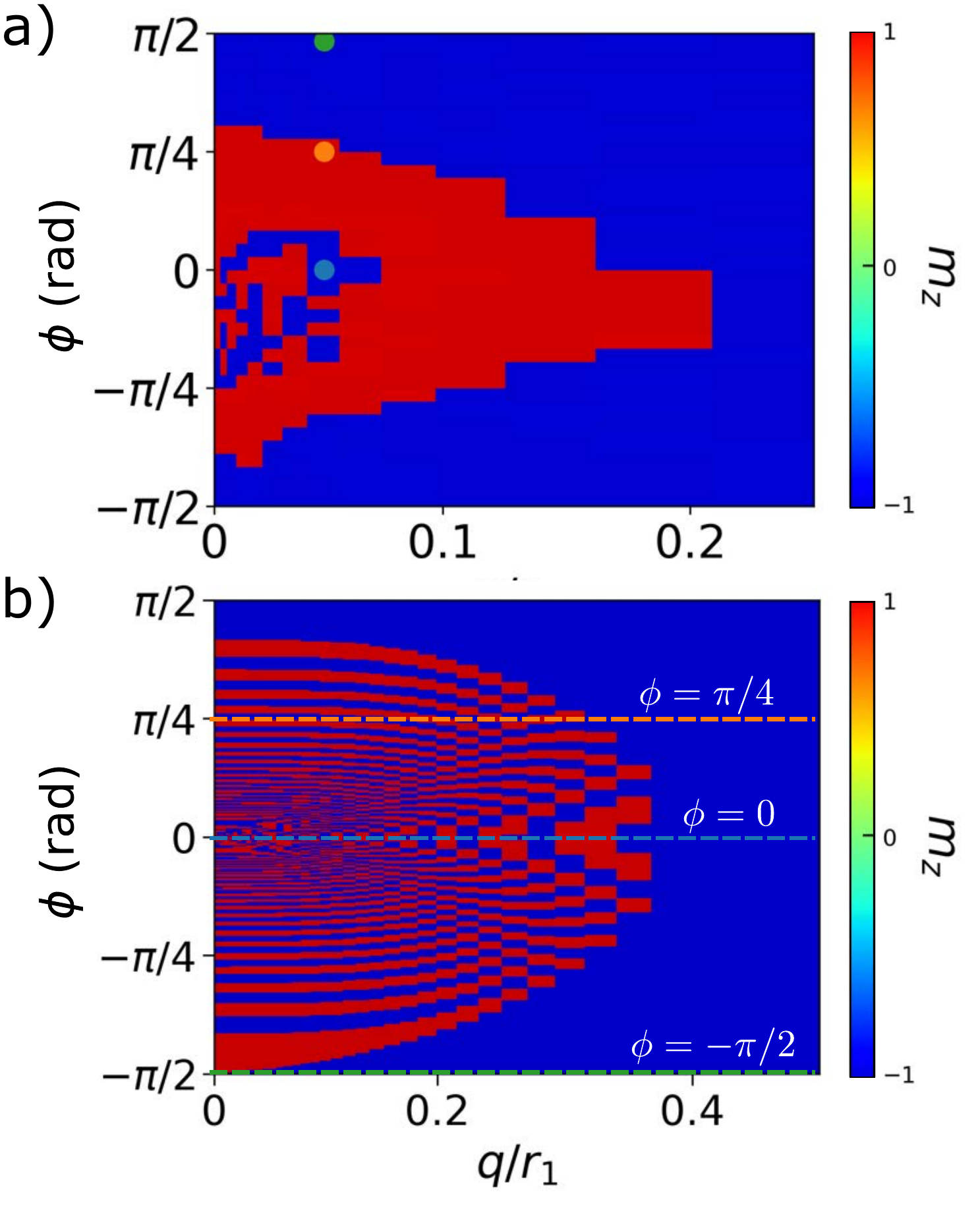}
     \caption{a) Relaxation phase diagram as a function of $q$ and $\phi$ with no spin current or field applied. We simulate a disk of radius $15$ nm for different initial domain sizes and different initial DW phases. The colorscale indicates the normalized magnetization of the final state where blue represents relaxation to a down domain and red that the magnetization switches. The blue, orange and green dots ($q/r_1=0.06$ and $\phi=0,\ \pi/4$ and $\pi/2$) show the parameters used in Fig.~\ref{FIG:relax_curves}. b) Relaxation phase diagram for the $(q,\phi)$ model. An asymmetry can be observed between positive and negative angles.}      
     \label{FIG:phase_diag}
\end{figure}

While there are differences in the timescales, the overall domain dynamics is captured by our analytic model. Importantly, they confirm that there are oscillations in the DW position and the phase runs, as in our analytic model. The micromagnetic simulations also corroborate the very sensitive dependence of domain dynamics on the initial conditions as the DW approaches the center of the disk. 

To examine this in more detail the relaxation dynamics were computed for a wide range of initial conditions and the results again used to construct a relaxation phase diagram. Figure~\ref{FIG:phase_diag}(a) shows this micromagnetic ($q,\phi$) phase diagram where we use the same colorscale as that in Fig.~\ref{FIG:plus_minus_q} and blue, orange and green dots represent the initial conditions for the curves in Fig.~\ref{FIG:relax_curves}. 

This pattern is compared to that of the collective coordinate model in Fig.~\ref{FIG:phase_diag}(b). The collective coordinate model is more intricate with many red/blue boundaries, showing a stronger dependence on initial conditions. Nonetheless, there are similarities. There is a region for small values of $q$ where the domain relaxes (blue region) and a red region in which switching is observed. In addition, the switching pattern is not mirror symmetric with respect to the initial phase. We attribute the asymmetry toward negative angles to the sense of gyroscopic motion. Depending on the initial phase, the DW will oppose or favor the rotation of the magnetic domain resulting in switching if the initial state favors the rotation.

\section{Dynamics driven by spin-polarized current} 
\label{Sec:SPC}
We now consider the influence of a spin current from a perpendicularly magnetized polarizing layer $\pvec=\hat{z}$ on the DW dynamics.
The spin-current threshold, $a_c$, 
needed to obtain a complete switch starting from a uniform state is related to the material parameters and can be derived from Eq.~\ref{Eq:LLGS} to be $a_c=\alpha \gamma \mu_0 H_k$, where $H_k$ is the perpendicular anisotropy field of the disk, $H_\mathrm{k}=2K_\mathrm{eff}/M$.
Our micromagnetic modeling shows that for parameters typical of state-of-the art magnetic tunnel junctions (see Appendix~\ref{Sec:AppendixPartE}) reversal by formation of a single domain wall for $1<d/d_c<2$~\cite{Mohammadi2020}. 
\begin{figure}[t]
  \centering
  \includegraphics[width=1\columnwidth]{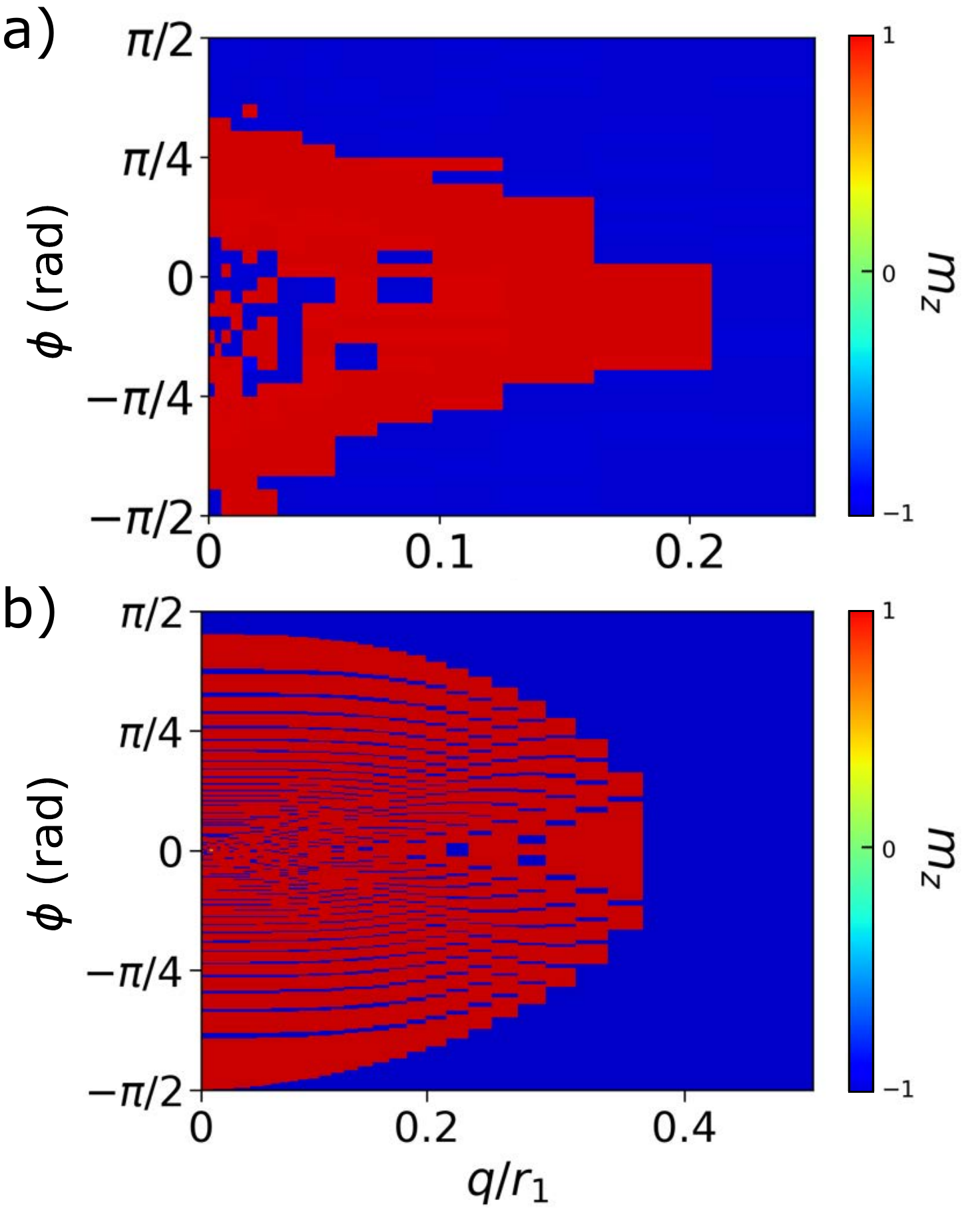}
     \caption{a) Relaxation phase diagram as a function of $q$ and $\phi$ for a spin current $a_I/a_c=0.05$, i.e. 5\% of the threshold current for switching starting from a uniform magnetization state. We simulate a $15$ nm radius disk for different initial DW positions and phases. The colorscale is the same as in Fig.~\ref{FIG:phase_diag}. b) Relaxation phase diagram for the parameters $q$ and $\phi$ calculated with the $(q,\phi)$-model for $a_I/a_c=0.07$.}
     \label{FIG:phase_diagram_10perI}
\end{figure}

In our analysis of the magnetic dynamics we consider that a single DW has already formed in the disk and model its relaxation, as in previous sections, but now in the presence of a spin-polarized current. Figure~\ref{FIG:phase_diagram_10perI}(a) shows a ($q,\phi$)-phase diagram calculated with micromagnetic simulations for the same parameters used in Fig.~\ref{FIG:phase_diag} with a spin current, $a_I/a_c=0.1$. As in the relaxation phase diagram, the final state seems to be sensitive to initial parameters. However, even in the presence of a small current (relative to the threshold current $a_c$), the parameters region resulting in a switched final state (red region) becomes bigger. This behavior is also captured by the analytic model in Fig.~\ref{FIG:phase_diagram_10perI}(b) where the effect of the current is more evident. A small spin current, $a_I/a_c=0.07$, increases the region with a switched final state (red region) and a current of $a_I/a_c=0.15$ eliminates the sensitivity to initial conditions and for $q/r_1\leq 0.4$ the dynamics results in switching.
\begin{figure}[t]
  \centering
  \includegraphics[width=1\columnwidth]{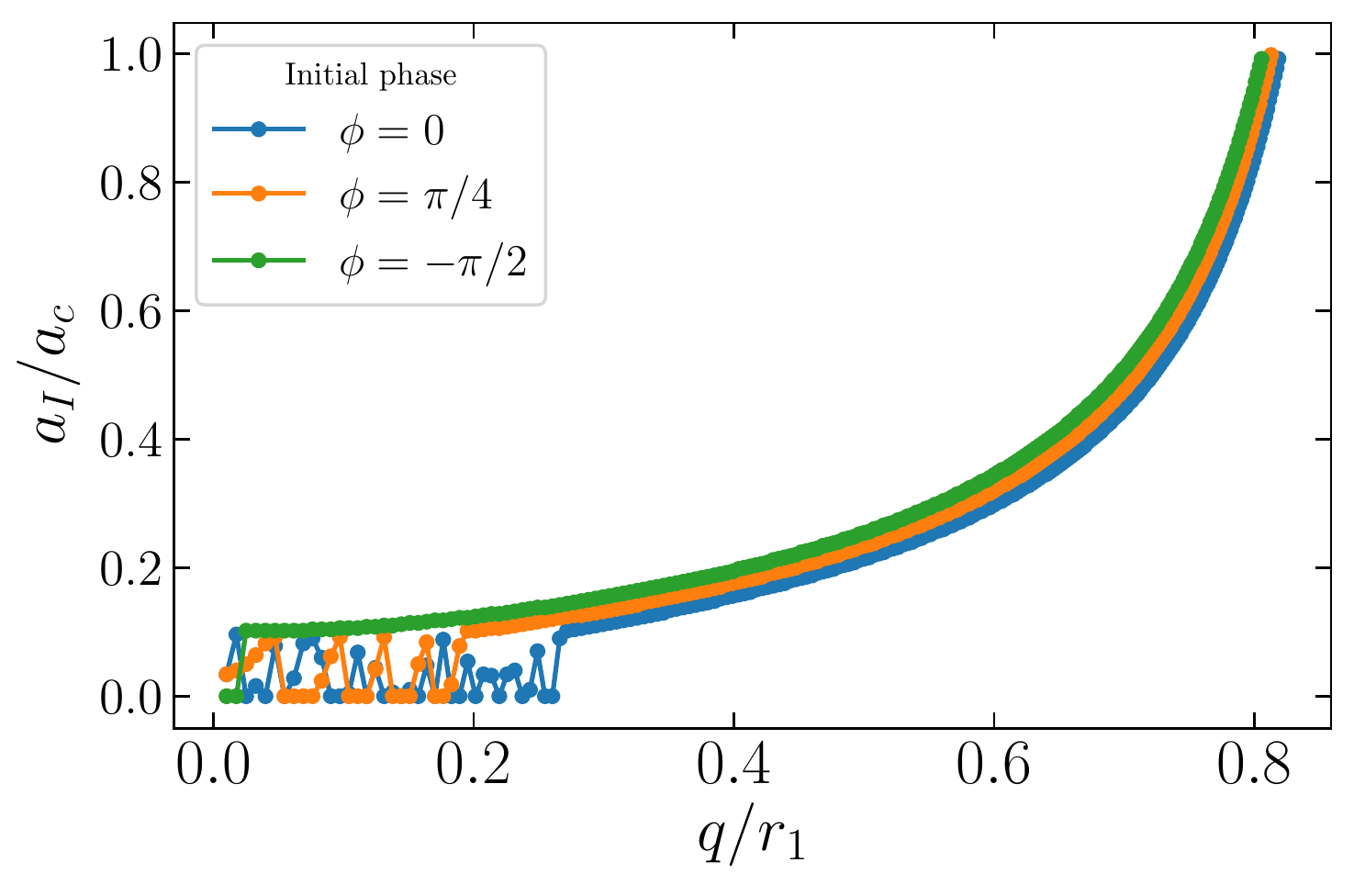}
     \caption{Switching current as a function of the initial position of the DW for three different initial phases. Each point indicates the minimum current required for magnetization switching.}
     \label{FIG:curr_vs_q}
\end{figure}

To more fully characterize the effect of spin current, we calculate the current needed for switching for different initial conditions. Figure~\ref{FIG:curr_vs_q} shows the current needed to switch the disk's magnetization as a function of the size of the initial domain for three fixed initial phases $\phi=0,\pi/4,-\pi/2$ (horizontal lines in Fig.~\ref{FIG:phase_diag}(b). We calculate the minimum current needed to switch the magnetization, which for some initial conditions is zero since the horizontal lines in Fig.~\ref{FIG:phase_diag}(b) cross a red area. For each initial phase there is a threshold for the size of the domain that separate the two different behaviors as a function of the current. For larger $q/r_1$, the switching current increases monotonically when the size of the domain is reduced, whereas for bigger domains---small $q/r_1$---regions can be found that switch  without current, which correspond with the red regions in Fig.~\ref{FIG:phase_diag}(b).
Similar behavior for the minimum current needed to switch the magnetization was observed in micromagnetic simulations that are shown in Appendix Sec.~\ref{Sec:AppendixPartE}. 

\section{Experimental consequences}
\label{Sec:EC}
The reversal mode and the micromagnetic instabilities will influence a pMTJ's electrical response to voltage pulses. This is because the junction conductance is related to the position of the DW. To a good approximation (neglecting the finite area of the DW) the junction conductance depends on the area of the reversed domain:
\begin{equation}
G=g_\mathrm{P} A_\uparrow + g_\mathrm{AP}  A_\downarrow,
\end{equation}
where $g_\mathrm{P} $ and $g_\mathrm{AP}$ are the specific junction conductances when it is magnetized in the parallel and antiparallel to the reference layer, respectively (units of $1/(\Omega$m$^2))$.
As the area of the disk (again, neglecting the finite area of the wall) is $A= A_\uparrow+A_\downarrow$:
\begin{equation}
G=(g_\mathrm{AP}-g_\mathrm{P})A_\downarrow+g_\mathrm{P}A.
\end{equation}
The last term $g_\mathrm{P} A$ is independent of time and since we are interested in the time dependence of the conductance we do not need to consider it further.

This shows that the conductance is proportional to area $A_\downarrow$. This area is in turn directly related to the normalized perpendicular magnetization of the disk,  $A_\downarrow = A(m_z/2+1/2)$. Therefore a measurement of the conductance versus time during switching would show the same behavior as $m_z$ versus time, e.g. that shown in Figs.~\ref{FIG:relax_bloch_phase_model} and~\ref{FIG:relax_curves}.

Direct imaging experiments may also be possible but are challenging at the length and time scales involved in these processes~\cite{Bernstein2011}. Single shot time resolved electrical studies would seem to be a promising means of observing these instabilities. The key model prediction is conductance oscillations that vary in frequency, increasing as the reversal proceeds to completion. In addition the model predicts that once a DW is nucleated lower spin currents can be used to displace it and reverse the magnetization.

\section{Summary}
In summary, we have considered the DW mediated magnetization switching of a disk in the presence of spin transfer torques; a geometry highly relevant to state-of-the art magnetic random access memory based perpendicular magnetized magnetic tunnel junctions. The results show a great sensitivity to initial conditions and, in particular, to the DW phase. An analytical model shows that DW surface tension leads to DW motion in the Walker breakdown limit. Key features of a simple collective coordinate model are found in micromagnetic simulations, including sensitivity to initial conditions and DW oscillations in the reversal process. These effects should be observable in experiments through measurements of tunnel junction conductance versus time. However, noise or other factors may modify the dynamics in real tunnel junctions. For example, noise can reduce the sensitivity to initial conditions and DW pinning associated with spatial variations in material parameters  (e.g. anisotropy, magnetization, etc.) will also effect the dynamics. These effects can be included in more realistic models that build on this research. 

\begin{acknowledgments}
We acknowledge useful discussions with Christian Hahn and Georg Wolf. This research was supported in part by National Science Foundation under Grant No. DMR-1610416. JMB was supported by Spin Memory Inc.
\end{acknowledgments}

\section{Appendix}
\subsection{Collective coordinate model}
\label{Sec:AppendixPartA}
An equation for the collective coordinates, $\xi_j$ and the generalized velocities $\dot{\xi}_j$, of a domain wall in disk can be derived from Eq.\ \ref{Eq:LLGS} following the approach outlined in Ref.~\cite{Tretiakov2008} 
\begin{equation}
G_{ij}\dot{\xi}_j+F_i-\Gamma_{ij}\dot{\xi}_j+S_\phi=0,
\label{Eq:Thiele}
\end{equation}
where:
\begin{eqnarray}
G_{ij}(\xivec)&=&J\int_\Omega{\mvec \cdot \left(\frac{\partial \mvec}{\partial \xi_i} \times \frac{\partial {\bf m}}{\partial \xi_j}\right) dV},
\label{Eq:Gij}\\
F_i(\xivec)&=&-\int_\Omega{\frac{\partial U}{\partial \xi_i}dV}=-\frac{\partial \mathcal{U}}{\partial \xi_i} \\
\Gamma_{ij}(\xi)&=&\alpha J \int_\Omega{\left( \frac{\partial \mvec}{\partial \xi_i}\right) \cdot\left(\frac{\partial \mvec}{\partial \xi_j}\right)}dV \\
S_i(\xivec)&=&a_I J \int_\Omega{\left( \frac{\partial \mvec}{\partial \xi_i}\right)\cdot (\mvec \times \pvec)}dV,
\label{Eq:Gamma}
\end{eqnarray}
where $\Omega$ is the region occupied by the ferromagnet, $J$ is the angular momentum density, $J=M/\gamma$ and $\mathcal{U}$ is the total energy of the system. The spin polarization is chosen to be perpendicular to the plane of the disk, $\textbf{p}=\hat{z}$, in what follows.
Substituting the magnetization texture of Eqs.~\ref{Eq:ms} into Eqs.~\ref{Eq:Gij}-\ref{Eq:Gamma} and integrating gives the coefficients in Eq.~\ref{Eq:Thiele} and resulting Eqs.~\ref{Eq:phi} and ~\ref{Eq:q} in the main text.

To solve Eqs.~\ref{Eq:phi} and ~\ref{Eq:q} numerically we rewrite them in the form: 
\begin{equation}
\frac{d{\bf y}}{dt}={\bf F}({\bf y},t),
\end{equation}
that is, we want the differential equations in a form in which there are no explicit time derivatives on the right hand side.

We thus rewrite Eqs.~\ref{Eq:phi} and ~\ref{Eq:q} as follows:
\begin{eqnarray}
(1+\alpha^2)\dot{\phi}&=&\frac{-\gamma}{2t \ell M} \frac{\partial \mathcal{U}}{\partial q}+\alpha \left[ \frac{\gamma \mu_0 H_N }{4}\sin{2\phi}-\frac{\pi a_I}{2}\right] \\
(1+\alpha^2)\dot{q}&=&\frac{\gamma \mu_0 H_N }{4}\sin{2\phi}-\frac{\pi a_I}{2}+\alpha \frac{\gamma}{2t \ell M} \frac{\partial \mathcal{U}}{\partial q}
\end{eqnarray}
To make the equations simpler we redefine the time as follows $\tilde{t}=t/(1+\alpha^2)$ and
\begin{equation}
\frac{d\phi}{d\tilde{t}}=(1+\alpha^2)\frac{d\phi}{dt}
\end{equation}
If we now write derivatives with respect to $\tilde{t}$ as $\phi'$ and $q'$ and the equations of motion become:
\begin{eqnarray}
\phi'&=&\frac{-\gamma}{2t \ell M} \frac{\partial \mathcal{U}}{\partial q}+\alpha \left[ \frac{\gamma \mu_0 H_N }{4}\sin{2\phi}-\frac{\pi a_I}{2}\right] 
\label{Eq:qdot1}\\
q'&=&\frac{\gamma \mu_0 H_N }{4}\sin{2\phi}-\frac{\pi a_I}{2}+\alpha \frac{\gamma}{2t \ell M} \frac{\partial \mathcal{U}}{\partial q}.
\label{Eq:qdot2}
\end{eqnarray}
Walker breakdown corresponds to $\phi'\ne 0$. The first term on the right hand side of Eq.~\ref{Eq:qdot2} indicates that the domain wall position $q$ oscillates when $\phi'\ne 0$. The condition for Walker breakdown, Eq.~\ref{Eq:Walker} in the main text, follows from Eq.~\ref{Eq:qdot1} with $a_I=0$.

Eq.~\ref{Eq:qdot2} shows that in the absence of a spin torque ($a_I=0$) and for magnetic fields less than the Walker breakdown field, the wall velocity is set by the phase $\phi$.
The maximum velocity occurs when $\phi=\pi/4$ and is $v_\mathrm{max}=\Delta \gamma \mu_0 H_N/4$.
This is the maximum velocity before Walker breakdown and we can write Eq.~\ref{Eq:qdot2}:
\begin{equation}
\frac{q'}{\Delta}= \frac{v_\mathrm{max}\sin{2\phi}}{\Delta}-a_I \pi/2+\alpha \frac{\gamma}{2t \ell M} \frac{\partial \mathcal{U}}{\partial q}.
\end{equation}

\subsection{Domain wall curvature}
\label{Sec:AppendixPartB}
     
As noted in Sec.~\ref{Sec:AM}, there is a term in the energy density associated with the curvature of the DW. This term comes from the exchange energy, the first term on the right hand side of Eq.~\ref{Eq:energydensity}. Including this term Eq.~\ref{Eq:totalE} becomes:
\begin{equation}
\mathcal{U}=\left[\sigma_\mathrm{DW}+\frac{\Delta \mu_0 H_N M}{2} \cos^2 \phi+ \frac{2 \Delta A}{r_2^2}\right] t \ell.
\label{Eq:totalE}
\end{equation}
This term leads to an infinite energy when $q=\pm r_1$, i.e. when the domain wall is at the boundary of the element. However, this energy only becomes important when the the DW is within its width $\Delta$ of the edge of this disk. In this case, the DW energy can decrease as the wall exits the disk. As result the DW energy is always finite and we can neglect this energy term in our analysis.

\subsection{Applied Field}
\label{Sec:AppendixPartC}
\begin{figure}[t]
  \centering
  \includegraphics[width=1\columnwidth]{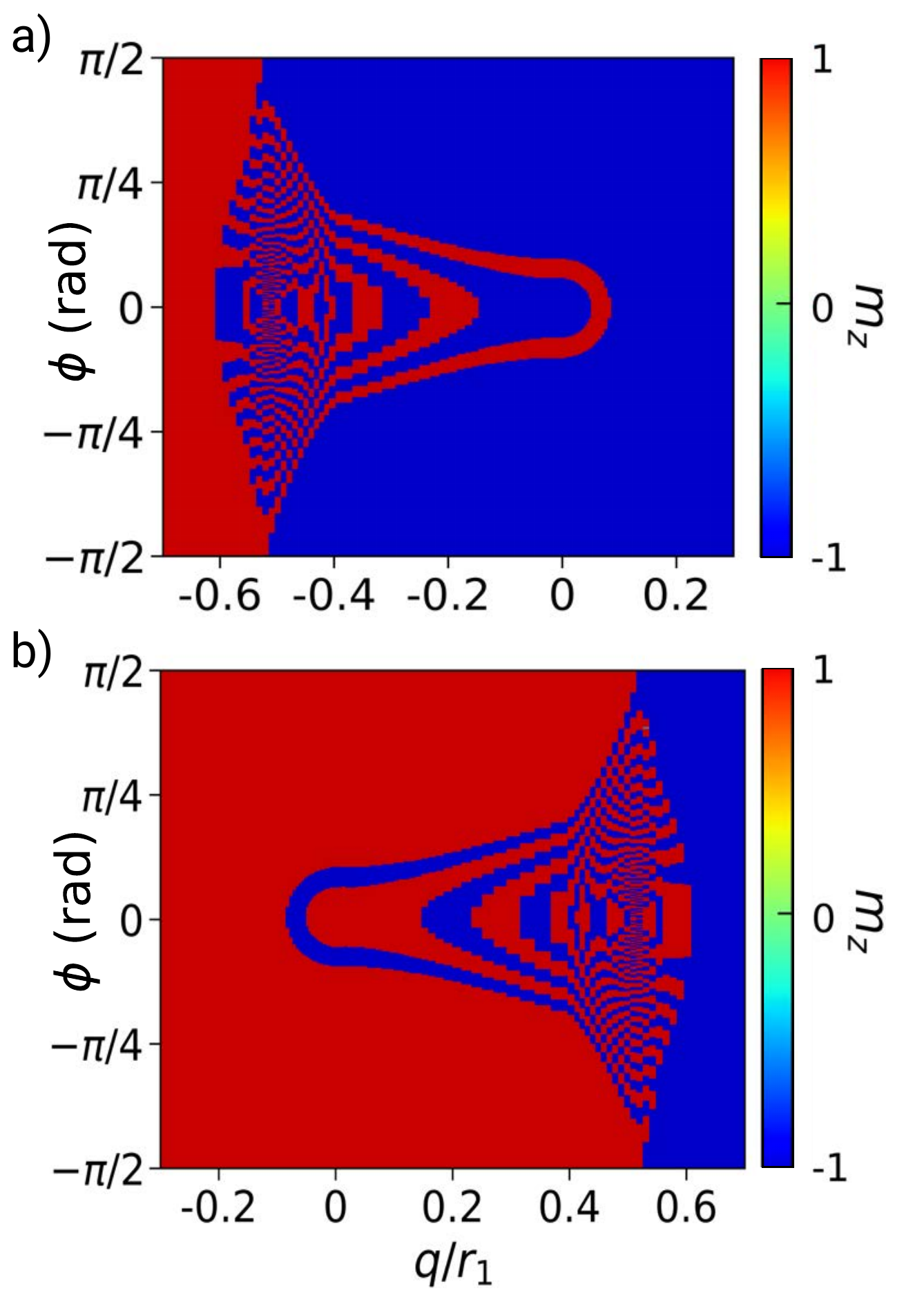}
     \caption{Relaxation phase diagram showing the final magnetization  state  as  a  function  of  the  initial  conditions  for  the analytic model in the presence of an field applied. (a) Field in the $-z$ direction and (b) field in the $+z$ direction. In both cases the applied field magnitude is half the coercive field $H=H_c/2$ and the spin current is zero. Red represents a final state with magnetization up and blue magnetization down.}
     \label{FIG:applied_field}
 \end{figure}

It is straightforward to include applied magnetic fields in the model. An applied field $H$ in the $z$ direction leads to the following set of equations of motion:
\begin{eqnarray}
\phi'&=&-\gamma \mu_0 H-\frac{\gamma}{2t \ell M} \frac{\partial \mathcal{U}}{\partial q}+\nonumber \\  && \alpha \left[ \frac{\gamma \mu_0 H_N }{4}\sin{2\phi}-\frac{\pi a_I}{2}\right] 
\label{Eq:qdot1H}\\
q'&=&\frac{\gamma \mu_0 H_N }{4}\sin{2\phi}-\frac{\pi a_I}{2}+\nonumber \\  && \alpha \left[\frac{\gamma}{2t \ell M} \frac{\partial \mathcal{U}}{\partial q}+\gamma \mu_0 H\right].
\label{Eq:qdot2H}
\end{eqnarray}
Figure~\ref{FIG:applied_field} shows the effect of the magnetic field on the final magnetization state as a function of the DW's initial conditions $(q,\phi)$. Figure~\ref{FIG:applied_field}(a) shows the results for a field applied in the negative $z$ direction and Fig.~\ref{FIG:applied_field}(b) shows the same diagram for a field applied in the opposite direction. In both cases the magnitude of the field is half the coercive field, $H=H_c/2$, where $H_c$ is defined as field that just renders the metastable magnetic state ($q=r_1$ or $q=-r_1$) unstable.

\subsection{Spin-polarized current with micromagnetics}
\label{Sec:AppendixPartD}
\begin{figure}[ht]
  \centering
  \includegraphics[width=1\columnwidth]{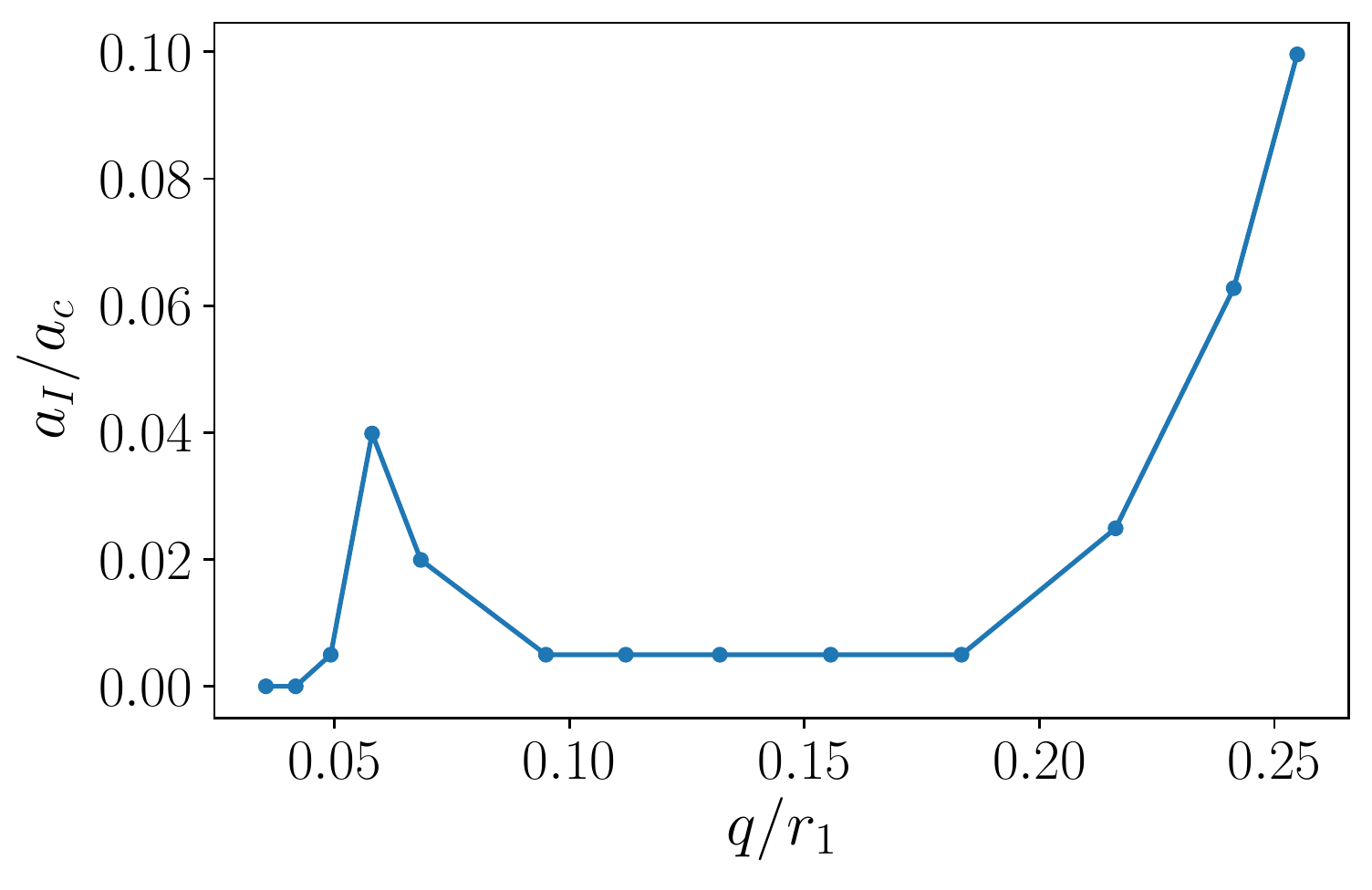}
     \caption{Switching current as a function of the initial position of the DW for $\phi= 0$ (Bloch DW) computed in micromagnetics. Each point indicates the minimum current required for magnetization switching.}
     \label{FIG:curr_vs_q_mumax}
\end{figure}

We also determined the current required for magnetization switching from different initial states with micromagnetics. Figure~\ref{FIG:curr_vs_q_mumax} shows the current needed to switch the disk's magnetization as a function of the initial DW position for $\phi= 0$. The behavior is qualitatively similar to that shown for the $(q,\phi)$ model in Fig.~\ref{FIG:curr_vs_q}, the current is small for small $q$ and increases as $q$ increases. The switching current is also a non-monotonic function of the DW's position $q$.

\subsection{Material parameters}
\label{Sec:AppendixPartE}
The parameters used in both the analytical model and the micromagnetic simulations were: saturation magnetization, $M=1.209\times 10^6$ A/m, damping constant $\alpha=0.03$, uniaxial anisotropy constant $K_p=1.12\times 10^6$ J/m$^3$ , exchange constant $A=4\times 10^{-12}$ J/m$^3$, a disk diameter of $30$ nm with thickness $t=2.3$ nm and demagnetization coeffecient $(3N_{zz}-1)/2\approx 0.74$~\cite{Chaves2015}. We performed micromagnetic simulations using the open-source MuMax$^3$ code \cite{Vansteenkiste2014} with a graphics card with 2048 processing cores. We considered the effects of Oersted fields but not finite temperature.
For a perpendicular magnetic tunnel junction, the zero-temperature critical current density~\cite{Sun2000} is related to materials parameters as follows:
$j_c=2e\alpha \mu_0 M H_k t/(\hbar P)$,
where $\alpha$ is the damping parameter, $e$ the charge of the electron, $H_k$ is the effective perpendicular anisotropy field that depends on the size of the junction~\cite{Chaves2015}, $t$ is the thickness of the disk and $P$ is the current polarization coefficient. For parameters we used in the calculations $j_c=9.23 \times 10^{10}$ A/m$^2$. We note that the characteristic field $\mu_0H_N=0.3$ T. The Walker breakdown field is $\mu_0 H_W=\alpha \mu_0 H_N/4=2$ mT.

\subsection{DW length and reversed domain area}
The following are some useful mathematical relations to compute the length of the DW and reversed domain area given the DW position. The angle betwen the disk's radius vector $r_1$ and DW position $q$ in Fig.~\ref{Fig:schematic_energy}(a) $\vartheta$ ($0\leq \vartheta \leq \pi$) is:
\begin{equation}
\vartheta=\arcsin{\left( \sgn{q} \cdot \frac{1-(q/r_1)^2}{1+(q/r_1)^2} \right)}.
\end{equation}
In terms of $\vartheta$ the length of the domain wall is $\ell=2r_1(\pi/2-\vartheta)\tan \vartheta$ and the area of the up magnetized domain is:
\begin{equation}
A_\uparrow/(\pi r^2)=\frac{\vartheta}{\pi}+\left(\frac{1}{2}-\frac{\vartheta}{\pi}\right)\tan^2\vartheta-\frac{\tan \vartheta}{\pi}.
\end{equation}


%

\end{document}